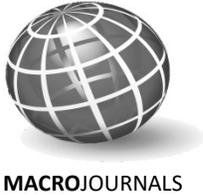

The Journal of **Macro**Trends in Technology and Innovation

# COMPARISON OF THE EFFICIENCY OF DIFFERENT ALGORITHMS ON RECOMMENDATION SYSTEM DESIGN: A CASE STUDY

**Gürkan Alpaslan**

*Department of the Computer Engineering, Yıldız Technical University, 34220, Istanbul, Turkey*

## Abstract

*By the growing trend of online shopping and e-commerce websites, recommendation systems have gained more importance in recent years in order to increase the sales ratios of companies. Different algorithms on recommendation systems are used and every one produce different results. Every algorithm on this area have positive and negative attributes. The purpose of the research is to test the different algorithms for choosing the best one according as structure of dataset and aims of developers. For this purpose, threshold and k-means based collaborative filtering and content-based filtering algorithms are utilized on the dataset contains 100\*73421 matrix length. What are the differences and effects of these different algorithms on the same dataset? What are the challenges of the algorithms? What criteria are more important in order to evaluate a recommendation systems? In the study, we answer these crucial problems with the case study.*

## I. INTRODUCTION

Recommendation systems provide suggestions to customers via evaluating their previous preferences [1] [2] or similarity of the topics or determining the similar users. The method of recommendation systems is filtering the huge data to extract the meaningful patterns [3]. Two main filtering methods are utilized for this purpose [4]: collaborative filtering [5] [6] and content-based filtering [7]. Collaborative filtering is the method that based on finding the similar users. The method assumed that the preferences of the similar users are relative with each other. Content-based filtering is the method that divides the data into little portions. The division is based on similarity of contents [8].

Recommendation systems have recently gained more importance due to increasing the e-commerce applications [9] [10]. To increase the profit, customers have to easily reach the topics they interest. For this purpose, as well as it is important to make over the results simultaneously depending on the customer's activities. As increasing the collected information about customer, the query results are considered more reliable.

In this research, a case study has been implemented on the dataset contains 100 articles and 73421 users. Articles have been evaluated by the users, but include the missing areas. On the dataset, every columns represent the articles and every rows represent the users. Figure 1 shows the structure of the data matrix.

$$
\begin{array}{ccccc}
 & a_1 & a_2 & \dots & a_n \\
u_1 & u_1a_1 & u_1a_2 & \dots & u_1a_n \\
u_2 & u_2a_1 & u_2a_2 & \dots & u_2a_n \\
\dots & \dots & \dots & \dots & \dots \\
u_n & u_na_1 & u_na_2 & \dots & u_na_n
\end{array}
$$

**Figure 1: The structure of data matrix**

In Figure 1, $u_1$ to $u_n$ represents the previous users; $a_1$ to $a_n$ represents the different articles and $u_xa_x$ represent the values of article x given by the user x. The values are between -10.0 to +10.0 numerical range. The dataset includes no nominal feature. Missing value range shows variability approximately between %18 to %70 depending on the article.

The main contribution of this work are: (1) comparing the different methods for recommendation systems, (2) proving the algorithm dependency for query combination and (3) demonstrating positive and negative ways of these algorithms.

The paper is structured as follows: in Section 2, the methods we utilized are described, in Section 3, we detail the case study, in Section 4, we describe the discussions and results, and finally, we draw some conclusion and future work.

## II.    MATERIAL AND METHODS

Three methods are implemented on the dataset:

- Threshold based collaborative filtering
- K-means based collaborative filtering
- Content-based filtering

For the threshold based collaborative filtering, a threshold value is firstly determined. The users who give points above the threshold are filtered. For these users, their other preferences are examined and provides to customer as suggestions.

K-means based collaborative filtering examines the rows like first method but main differences is filtering procedure of the similar users. The dataset includes no nominal feature. In this method, firstly whole dataset classifies as class 0 to class 99 using k-means classification. K-means classification is a method for determining the closeness of the points and classify. K is the number how many subsets algorithm divide the dataset. After that, via utilizing the current preferences of the referenced customer, its class is determined, so that, other 99 classes are ignored. For the determined class, popular articles are suggested.

Content-based filtering divides the articles into subsets via similarity of their mean values. In this method, the columns of the dataset are examined in contradistinction to other two methods. Mean values of every columns are determined and the columns between –x to +x range near to that are determined as same subset elements. The value of x are determining according to distribution of the dataset features. After classification, the preferences are evaluated only in their classes.

## III. CASE STUDY

An implementation that runs the algorithms has been coded.

The system working principle is designed as follows:
- Customer requests the articles
- Articles are shown randomly
- Customer chooses the articles he likes
- The articles he likes compose the transaction that will be queried
- For this study, transaction length is limited as four elements
- After customer chose the four articles, system requires selecting the filtering method
- As the selection, the results are given to customers.

The every algorithm runs 1000 times and the results are evaluated in criteria: run-time performance, query dependency and result set length.

## IV. RESULTS AND DISCUSSION

Threshold based collaborative filtering algorithm runs depending on the transaction query. As the query combination, it gives no results or lots of results more than expected. It has high query-dependency.

K-means based collaborative filtering algorithm runs the using of Weka [11] libraries. Algorithm runs slowly, due to the data clustering time. It has low performance, on the other hand, algorithm gives results for all queries. It is query-independent.

Content based-filtering runs fast and has little query dependency. It runs depending on the distribution of the features of the articles.

**Table 1. Comparison of the Algorithms**

|  | Run-time Performance | Query Dependency |
|---|---|---|
| Threshold based algorithm | Medium | High |
| K-means based algorithm | Low | Low |
| Content based algorithm | High | Medium |

By content based algorithm, we have run the algorithm repeatedly 1000 times. Table 2 shown the recommendation set length for four selected articles as input.

**Table 2. Result set length for Content-based Algorithm**

| Output length | Empty set | 1 | 2-3 | 4-9 | 10> | Sum |
|---|---|---|---|---|---|---|
| Amount | 1 | 8 | 80 | 782 | 129 | 1000 |
| Proportion | %0.1 | %0.8 | %8 | %78.2 | %12.9 | %100 |

As shown in Table 2, the algorithm has been returned empty set only 1 time in 1000 iterations. It evaluated very bad result and its proportion is %0.1. Best proper length is assumed 4 to 9 length and its proportion is %78.2.

For threshold based content based filtering algorithm, 1000 iterations have been implemented. For this algorithm, the threshold value is the critical factor for effectiveness. For testing, determined threshold value is 9.3 and under these conditions, result set is evaluated on the criteria: empty set, very short, short, normal and long; since the result set length are taken value between 0 to 47 so it has been normalized depends on the density distribution.

**Table 3. Result set length for threshold based collaborative filtering algorithm**

| Output length | Empty set | Very Short | Short | Normal | Long | Sum |
|---|---|---|---|---|---|---|
| Amount | 8 | 16 | 54 | 824 | 98 | 1000 |
| Proportion | %0.8 | %1.6 | %5.4 | %82.4 | %9.8 | %100 |

K-means based collaborative filtering algorithm is independence for result set length. Algorithm takes the best options of every cluster even the points of the options are low. So, algorithm give the result for every condition and it give the opportunity to determine the result set length.

## V. CONCLUSION

In the study, three algorithms have implemented for creating a recommendation system on the same dataset. The study obviously show the crucial effect of the criteria: query-dependency and run-time performance on recommendation systems for accuracy, usability and efficiency. As well as, we can obviously conclude that the length of the result set is crucial point for recommendation systems.

As the algorithms run a thousand of times, though a very low likelihood, algorithm has been concluded the query as empty set. This situation is another crucial point, even it occurs in low possibility. Meanwhile, k-means based collaborative filtering algorithm gives result for any input query combination; so it outshines thanks to this advantage.